\documentclass[%
reprint,
showpacs,preprintnumbers,
amsmath,amssymb,
aps,
prb,
]{revtex4-1}

\usepackage{epstopdf}
\usepackage{graphicx}
\usepackage{dcolumn}
\usepackage{bm}
\usepackage{hyperref}
\usepackage{ulem}
\usepackage{color}

\begin{document}

\title{Effect of charging on CdSe/CdS dot-in-rods single photon emission.}%

\author{M. Manceau$^{1}$, S. Vezzoli$^4$, Q. Glorieux$^1$, F.Pisanello$^2$,
E.Giacobino$^1$, L. Carbone$^3$, M. De Vittorio$^{2,3}$ and A. Bramati$^1$}

\affiliation{
 1 Laboratoire Kastler Brossel,Universit\' e Pierre et Marie Curie, CNRS UMR
8552, Ecole Normale Supérieure,4 place Jussieu, 75252 Paris Cedex 05, France\\
 2 Istituto Italiano di Tecnologia (IIT) Center for Bio-Molecular
Nanotechnologies Via Barsanti sn, 73010 Arnesano (Lecce), Italy\\
 3 National Nanotechnology Laboratory (NNL)- CNR Istituto Nanoscienze, c/o
Università del Salento, via per Arnesano km 5, 73100 Lecce, Italy\\
 4 Center for Disruptive Photonic Technology (CDPT)
School of Physical and Mathematical Sciences (SPMS)
Nanyang Technological University
}

\date{\today}

\begin{abstract}

The photon statistics of CdSe/CdS dot-in-rods nanocrystals is
studied with a method involving post-selection of the photon
detection events based on the photoluminescence count rate. We show that
flickering between two states needs to be taken into account to
interpret the single-photon emission properties.  With
post-selection we are able to identify two emitting states: the exciton 
and the charged exciton (trion),
characterized by different lifetimes and different second order correlation functions.
Measurements of the second order autocorrelation function at zero delay with post-selection shows 
a degradation of the single photon emission for CdSe/CdS dot-in-rods
in a charged state that we explain by deriving the neutral and charged biexciton 
quantum yields.

\end{abstract}

\maketitle

\section{Introduction}\label{sec:Introduction}

Wet-chemically synthesized colloidal core/shell nanocrystals (NCs)
have been shown to emit non-classical light at room temperature
\cite{michler2000,lounis2000photon,brokmann2004colloidal,fisher2005room}
making them suitable for applications in the field of quantum
optics. Among colloidal NCs, CdSe/CdS dot-in-rods (DRs) made up by a 
spherical core of CdSe surrounded by a rod-like
shell of CdS\cite{carbone2007synthesis} have interesting features
as single photon emitters such as a high degree of linear
polarization \cite{vezzoli2013ensemble,pisanello2010room,sitt2011highly,hadar2013polarization,pisanello2010dots} 
and a strongly reduced emission intermittency (non-blinking emission)
when synthesized with thick CdS shells \cite{pisanello2013}. The emission of these
nanostuctures at room temperature comes from the relaxation of various
states, namely the exciton, the charged exciton and the
neutral and charged multiexcitons. This diversity of emission states
makes the interpretation of photon
statistics quite difficult. Different emission states are
characterized by different photons statistics and a study
of the whole photons set detected from a given nanocrystal only gives an
average behavior which can differ from particle to particle in the same batch. 
The photon statistics study is also a way to retrieve 
information about the charges relaxation processes such as Auger non-radiative recombination
\cite{park2011near,zhao2012biexciton}. A deeper understanding of the photon statistics
is therefore a requirement for the comprehension of the emission processes of such nanoemitters.

Auger process in colloidal nanocrystals is at the origin of
single-photon emission \cite{klimov2000quantization,fisher2005room}
and it is suggested to be involved in blinking process
\cite{nirmal1996fluorescence,efros1997random,pisanello2013} in such structures.
Auger effect is a three-charge process in which the recombination
energy of an electron-hole pair is transferred to an additional charge. 
Fast Auger non-radiative recombinations of
multiexciton states leads to a nonradiative cascade decay of the
excited states explaining the single-photon emission, hence multiphoton
emission is quenched. However, if an extra charge is present in
the crystal, Auger non-radiative recombination with the extra charge leads to the
quenching of the exciton emission and blinking. Recent works
\cite{spinicelli2009bright,galland2011two,galland2012lifetime} on
CdSe/CdS spherical dots have demonstrated the importance of the
geometry and the charge delocalization in influencing the Auger recombination rates.
By growing NCs with a thicker shell around the core, it has been shown that
charged nanocrystals still emit light  owing to a less efficient
Auger transfer. A negatively charged nanocrystal in a trion state
has been shown to have  an emission
efficiency lower \cite{spinicelli2009bright,galland2011two}
than or equal \cite{galland2012lifetime} 
to the pure exciton state. Several works have been devoted
to the study of photon statistics as a method to assess the
efficiency of the Auger processes and their consequence on
multiexciton recombinations and
blinking\cite{park2011near,zhao2012biexciton}. Nair et al.
\cite{nair2011biexciton} have demonstrated that one 
can deduce the biexciton quantum yield from the intensity
autocorrelation function at zero delay, $g^{(2)}(0)$, for NCs pumped
at low fluences.

In the present paper, we study the emission properties of DRs in
terms of photon statistics. We show that a degradation of the single photon emission is 
associated with a charged nanocrystal by post-selecting photons based on their associated 
count rates after binning the signal \cite{spinicelli2009bright}. This method allows to quantify separately the photon
statistics associated with the two states of charge of such nanostructures.  
We explain the difference of photon statistics between a neutral and charged DR by evaluating 
the neutral and charged exciton and biexciton quantum yields.
First, we present the emission characteristics of our DRs
in terms of a flickering between a bright and a grey state. Thanks
to post-selection on the photon detection events, we can reconstruct
the autocorrelation function of the bright and grey state
separately.  Comparing two DRs from the same batch, which display different exciton lifetimes,
we explain the differences observed for the photon statistics in terms of charge delocalization and
flickering between the two states.

\section{Experimental set-up}\label{sec:Experiment}

\begin{figure*}
  \centering
\includegraphics[scale=1]{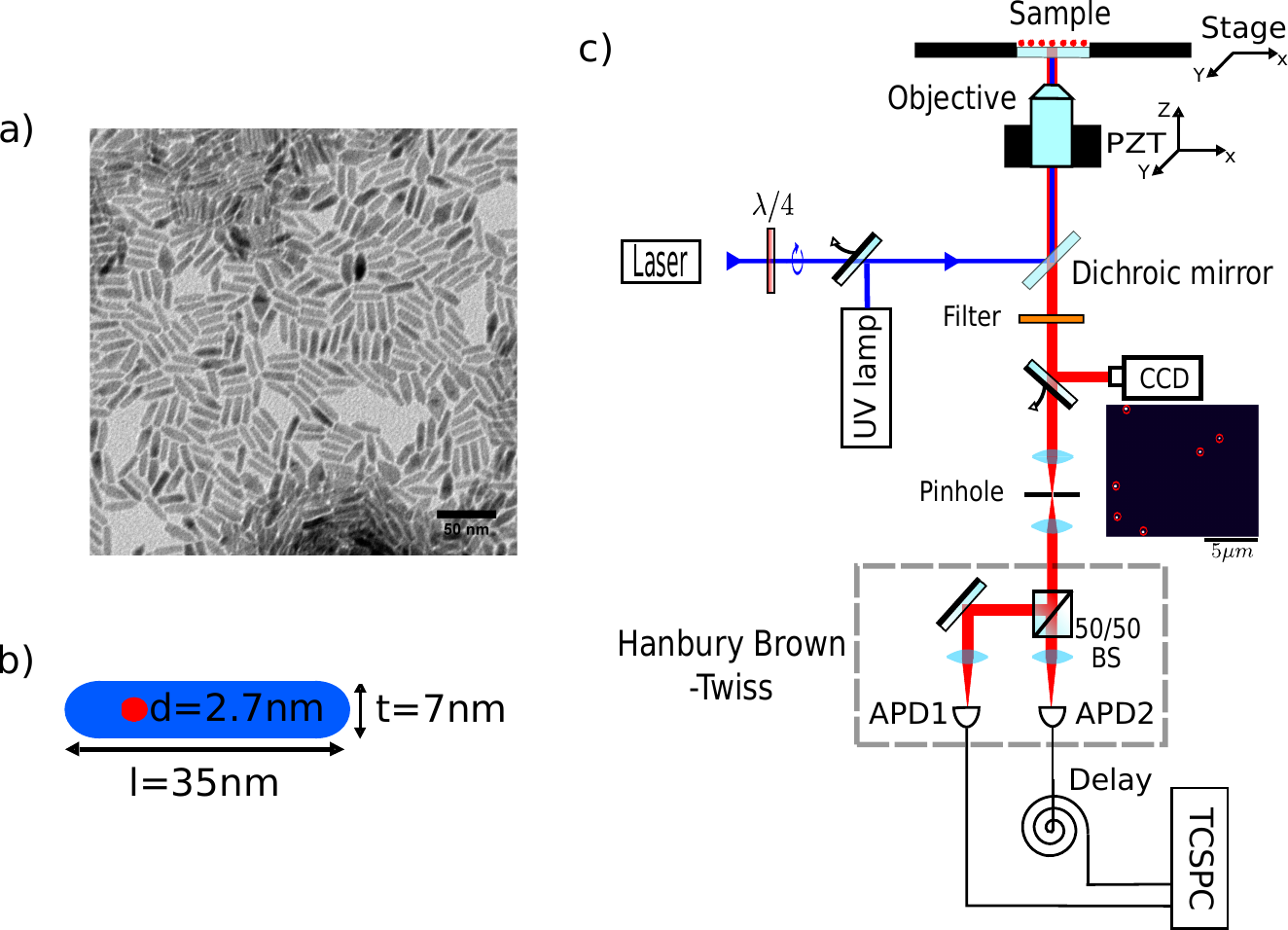}
  \caption{\textbf{a)}Transmission electron microscope (TEM) image of the investigated DRs before
  dilution.
 \textbf{b)}DR geometrical features determined from statistical analysis 
 performed on TEM images. Shell length $l=35$~nm, shell
thickness $t=7$~nm and core diameter $d=2.7$~nm.
 \textbf{c)}Experimental setup.  
Wide field microscopy can be realized by exciting the sample with a UV lamp. In this case, a CCD camera is used to image the excited area 
of the sample. A zoom on a typical CCD camera image is presented with DRs photoluminescence highlighted by a red circle.
Circularly polarized light from a picosecond-pulsed excitation laser diode with a wavelength of
$404$~nm, a pulse width of $100$~ps and a repetition rate of $2.5$~MHz is
used to excite single DRs. A confocal microscope collects the photoluminescence of the DRs. A high pass filter (cutoff $570$~nm) removes 
the remaining blue light from the optical path. Single photons events are recorded by two avalanche photodiodes in a Hanbury-Brown and Twiss
configuration. $\lambda/4$: Quarter wave plate. BS: Beam Splitter. APD: Avalanche photodiode.
TCSPC: Time-Correlated Single Photon Counting.}
  \label{fig:Setup}
\end{figure*}

We used high quality CdSe/CdS core-shell DRs synthesized  using the
seeded growth approach proposed in reference\cite{carbone2007synthesis,pisanello2013}. A transmission 
electron microscope (TEM) image of the investigated sample before dilution is presented in Fig.\ref{fig:Setup}a. 
The DRs are characterized by a shell length $l=35$~nm, a thickness $t=7$~nm and a core diameter $d=2.7$~nm
as depicted in Fig.\ref{fig:Setup}b. 
A dilute solution is drop-casted on a microscope glass coverslip to produce a
low density of single DRs ($2$ to $5$ DRs per $5$~$\mu$m$^2$ area).
Wide field luminescence microscopy can be realized using a UV lamp to excite a broad area of the sample.
Imaging of the excited area on a high quantum efficiency CCD camera gives an overview of the sample as shown in
Fig.\ref{fig:Setup}c with a zoom on a $12$~$\mu$m$^2$ area.
A single DR can be subsequently chosen and excited using a picosecond-pulsed laser diode with a small excitation 
spot of $1$~$\mu$m$^2$. The picosecond-pulsed laser operates at a wavelength of $404$~nm to excite the highly
absorptive
shell \cite{carbone2007synthesis}, with a repetition rate of $2.5$~MHz such that the time between two laser pulses is
typically greater than an absorption emission cycle duration. 
To obtain an excitation independent of DRs orientations on the substrate the laser light is circularly polarized.
The photoluminescence (PL) is collected using a confocal microscope
with a high numerical aperture oil immersion objective ($100 \times$, N.A.=$1.4$). A high pass filter (cutoff
$570$~nm) removes the remaining excitation light while leaving the DRs photoluminescence which is centered around $600$ nm. The DR photoluminescence is 
then spatially filtered through a pinhole (diameter $150$~$\mu$m) and subsequently recorded using two single-photon avalanche
photodiodes (APD)  in a Hanbury-Brown and Twiss configuration as shown in Fig.\ref{fig:Setup}c.  
The signals from the photodiodes were recorded by a Time-Correlated Single Photon
Counting (TCSPC) data acquisition card (PicoHarp300,Picoquant) enabling for each DR the recording of the PL lifetimes
and the PL autocorrelation function.

\section{Flickering}\label{sec:Flickering}

The DRs investigated in this study are well described by a fast
switching (flickering) between two states of emission: a bright
state with a high emission efficiency and a grey state with a lower
emission efficiency \cite{spinicelli2009bright,pisanello2013}. The emission efficiency
is assessed by the quantum yield (QY) which is given by the ratio
between the radiative decay rate and the total decay rate due to
radiative and non-radiative relaxations. Excitons in high quality
CdSe structures have a QY close to unity
\cite{brokmann2004measurement,fisher2004emission}, while the trion
state exhibits a QY ranging from $15$ to $50\%$
in this specific sample \cite{pisanello2013}.

\begin{figure*}
  \centering
\includegraphics[scale=0.5]{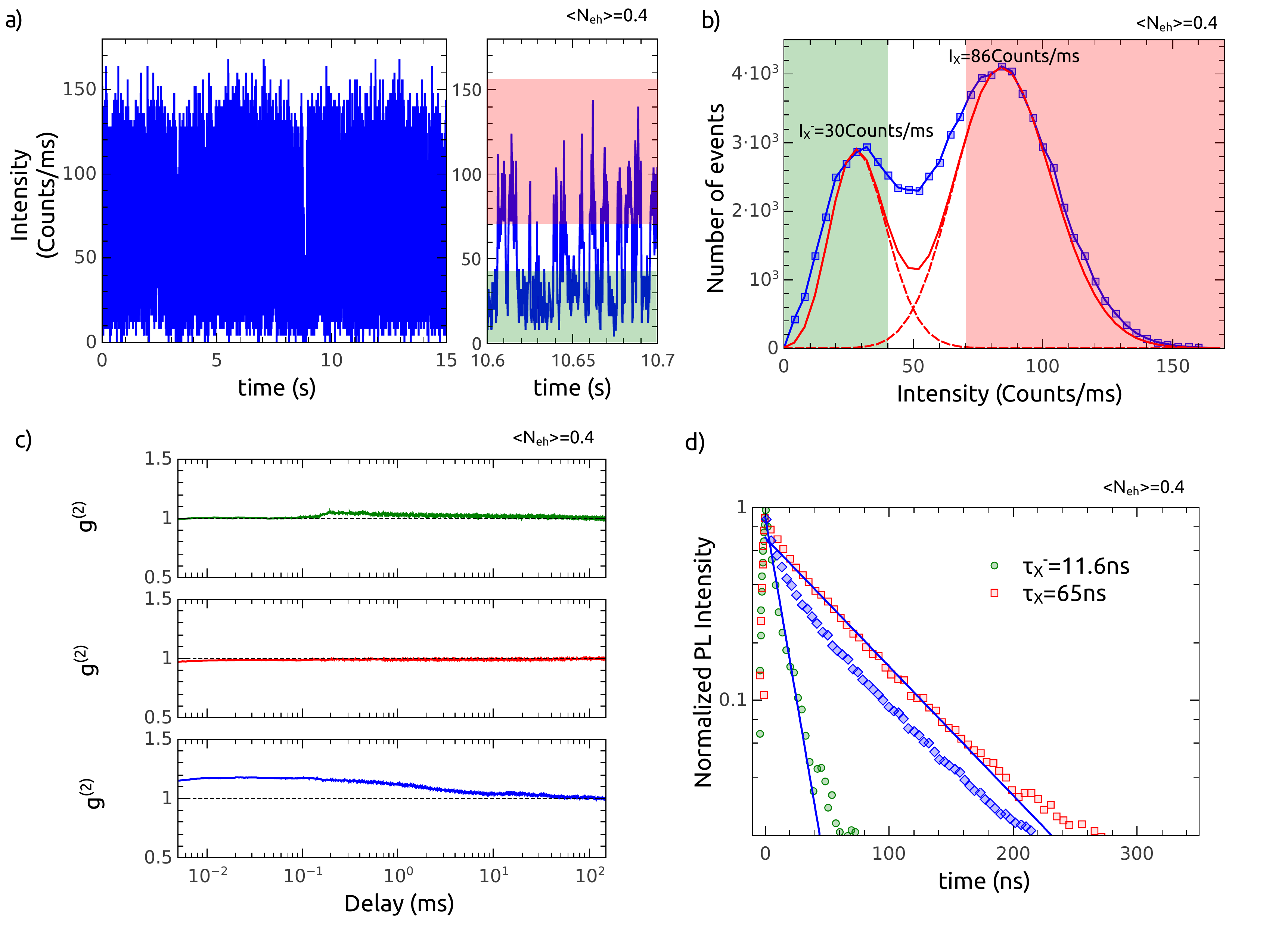}
  \caption{Study of DR1 photoluminescence.  \textbf{a)} PL intensity of DR1 on second (left)
and millisecond (right) timescales for a mean average
number of excitons in the nanocrystal of $\langle Neh \rangle=0.4$
and a bin time of $250$ $\mu$s.
  \textbf{b)} Histogram of the PL intensity shown in a) (number of events corresponding to a given
intensity per time bin) together with a fit with two Poissonian
distributions.  The bright state in red has a mean
emission intensity $I_X=86$~counts/ms,
the grey state in green has a mean emission intensity
$I_{X^{-}}=30$~counts/ms.
  \textbf{c)}$g^{(2)}$ of the PL intensity on
several decades of delays between photons. Top: $g^{(2)}$ of the grey state (green
intensity window in a and b). Middle: $g^{(2)}$ of the bright state (red intensity
window in a and b)
  Bottom: $g^{(2)}$ of the whole PL intensity.
  \textbf{d)} Decay of the PL intensity. Blue diamonds: PL decay of the whole intensity distribution in b).
Red squares: PL decay for the photons in the red intensity window in b), a
monoexponential decay fit
  gives $\tau_X=65$~ns. Green circles: PL decay for the photons in the green
intensity window in b),  a monoexponential decay fit gives $\tau_{X^-}=11.6$ns.}
  \label{fig:Lifetimes}
\end{figure*}

Representative emission properties from a single DR, namely DR1, are displayed in
Fig.\ref{fig:Lifetimes}.
Fig.\ref{fig:Lifetimes}a shows the photoluminescence (PL)
intensity recorded during 15 seconds with a bin time of
$250$~$\mu$s, for a low excitation power corresponding to an average
number of electron hole pairs excited by a single laser pulse of $\langle Neh
\rangle=0.4$. A zoom on a $100$~ms time window is also shown. The switching between 
the two states is clear on the shorter time window of $100$~ms. The
red part of the time trace above $70$~counts/ms corresponds to the
bright state while the green part below $40$~counts/ms can be
attributed to the grey state. In Fig.\ref{fig:Lifetimes}b, the
histogram of the PL intensity confirms the presence of two states. A
fit of this histogram with two Poissonian distributions reproduces
the intensity distribution properly, except in the intermediate
region. In the following, the post-selected photons of a given state are chosen 
based on the intensity histogram. If a time bin has a number of photons such that 
it falls into a state intensity window, then the photons of this time bin are 
associated to this state. All subsequent data analysis for a given state, such as PL lifetime or 
autocorrelation function, are realized with the chosen photons from the intensity histogram.

For long timescales, Fig.\ref{fig:Lifetimes}c shows the normalized second order autocorrelation
functions (ACF), $g^{(2)}$ as given by Eq.(\ref{g2}) at the
beginning of section \ref{sec:Flickering_and_Biexciton_emission}. Long timescales here means that 
we present the $g^{(2)}$ function with a resolution and on timescales longer than the emitter lifetime and laser
repetition rate. Therefore no antibunching and quantum properties of the emitter are revealed with this measurement
but information on the flickering can be retrieved.
From top to bottom respectively, the $g^{(2)}$ for the photons inside the grey state count rate range
($0$ to $40$ counts/ms), the $g^{(2)}$ for the photons inside the bright state count rate
range (above $70$ counts/ms) and finally the $g^{(2)}$ of all the
recorded events displayed in Fig.\ref{fig:Lifetimes}a are shown. When no count rate range is chosen
(Fig.\ref{fig:Lifetimes}c bottom), a super-Poissonian statistics
with $g^{(2)}$ of $1.2$ is observed on timescales ranging from
$1$ $\mu$s to $10$~ms due to the flickering between the two states
\cite{fleury2000nonclassical,messin2001bunching}. For delays above
$10$~ms the $g^{(2)}$ falls down to one, meaning that switching
between the two states does not happen on these longer timescales
owing to a reduced blinking for these thick shells DRs
\cite{pisanello2013}. As flickering occurs even at microseconds timescales,
it is always faster than any typical bin time used to build the
intensity distribution. There is no perfectly appropriate bin time
to fully discriminate the two states and binning the signal always
leads to a mix of the two states. Therefore the transition between the
two states is blurred and the system is not  well
described by a superposition of two Poissonian states in the
transition region as seen in Fig.\ref{fig:Lifetimes}b (blue line). In contrast,
the $g^{(2)}$ functions in Fig.\ref{fig:Lifetimes}c for the selected
photons of the grey (top) and bright (middle) states have a value of
one at every timescales displayed. This is a strong evidence that our
post-selection of the photon detection events discriminates well
between the photons of the two states  as no additional bunching
(super-Poissonian statistics) due to the flickering is present. The
photons associated with the intensities between  $40$~counts/ms and
$70$~counts/ms inside the central region of the intensity histogram
in Fig.\ref{fig:Lifetimes}b cannot be attributed to a specific state 
and have therefore been discarded. They represent $17\%$ of the 
photon events recorded for this specific measurement, while the grey and
bright state photons represent $6.5\%$ and $76.5\%$ of the recorded photons 
respectively.

We further confirm the analysis of the PL properties of our DRs in
terms of switching between a bright and a grey state by studying the
PL decay. In Fig.\ref{fig:Lifetimes}d, while the full PL decay (blue
diamonds) cannot be fitted with a single exponential, post-selecting
only the photons recorded in the red intensity window in
Fig.\ref{fig:Lifetimes}d (red squares), the PL decay is nearly
mono-exponential with a decay constant of $\tau_{X}=65$~ns.
 The PL of the photons in the green intensity
window (green circles) in Fig.\ref{fig:Lifetimes}c is found
to decay exponentially with a time constant $\tau_{X^-}=11.6$~ns.
These two time constants are attributed to the neutral and the
negatively charged exciton (X and $X^-$) as demonstrated in the literature\cite{galland2011two}.
They correspond to the bright and grey states respectively.
These results allow to evaluate the emission quantum
yields. The exciton quantum yield is defined as:
$Q_{X}=\gamma_r/(\gamma_r+\gamma_{nr})=\gamma_r\tau_{X},$
with $\gamma_r$, $\gamma_{nr}$ the radiative and non-radiative decay
rates respectively and $\tau_X$ the measured exciton lifetime.
Assuming $Q_{X} \simeq 1$ \cite{brokmann2004measurement,fisher2004emission,pisanello2013}, 
the radiative decay rate is simply the inverse of the measured exciton lifetime.
Using statistical scaling
\cite{klimov2008scaling,galland2012lifetime}, the trion
radiative decay rate can be shown to be twice as fast as
the neutral exciton because of the creation of a new radiative
relaxation path by the extra charge (an electron in this case). The trion QY is therefore given by:
$Q_{X^{-}}=2\gamma_r/(2\gamma_r+\gamma_{nr})=2\gamma_r\tau_{X^{-}}=2\tau_{X^{-}
}/\tau_X$, which gives
$Q_{X^{-}}=2\times 11.6/65=36\% $, using the data of Fig.\ref{fig:Lifetimes}d. The mean intensities in
Fig.\ref{fig:Lifetimes}.b where $I_{X^-}=30$~counts/ms is equal to
$35\%$  of  the mean exciton intensity ($I_{X}=86$~counts/ms)
confirms the statistical scaling approach and the consistence of our
model. The lower QY of the trion comes from the fact that the extra electron 
opens not only a new radiative relaxation channel but
also a non-radiative channel. This non-radiative channel is due to
an Auger process\cite{klimov2008scaling}, which is the relaxation
of the electron hole pair to the extra electron. The non-radiative
decay rate $\gamma_{nr}$ for a negatively charged DR is therefore
equal to the  Auger relaxation rate: $\gamma_{nr}=\gamma_{A^{-}}$.

We have shown that the emission of our DRs is well described by a
fast switching between a bright and a grey state. Post-selection
of the photon detection events based on the intensity count rates
for a given bin time can successfully
discriminate between the photons of each state as attested by the
$g^{(2)}$ functions in Fig.\ref{fig:Lifetimes}c and by the pure
exponential decay of the two states. In the following, using
post-selection of the photon events, we show that we can retrieve
the biexciton and charged biexciton QYs from the autocorrelation
function at zero delay $g^{(2)}(0)$ and explain the differences in
the photon statistics obtained with different DRs.

\section{Flickering and Biexciton
emission}\label{sec:Flickering_and_Biexciton_emission}

The normalized intensity autocorrelation function $g^{(2)}(\tau)$ is obtained
from the numbers of counts  $I_{1}$ and $I_{2}$ measured in the two
channels of the Hanbury-Brown and Twiss set-up as a function of the
delay $\tau$ between the two channels. It is given by:

\begin{equation}\label{g2}
    g^{(2)}(\tau)= \frac{\langle I_{1}(t)I_{2}(t+\tau) \rangle} {\langle
I_{1}(t)\rangle\langle I_{2}(t+\tau)\rangle}.
\end{equation}

\noindent In ref.\cite{nair2011biexciton}, a general equation is
derived to express the autocorrelation function at zero delay for a
single nanocrystal depending on the probability to excite a given
number of electron-hole pairs and the multiexciton quantum yields of
the different excited states. The probabilities $P_{N_{eh} \geq m}$
that at least $m$ excitons are created inside the nanocrystal for an
excitation energy well above the exciton line, in the continuum of
the shell states, a condition typically realized in our experiment,
is:

\begin{equation}\label{PoissonianProb}
P_{N_{eh} \geq m} = \sum_{m^{'}\geq m}^{}\frac{{\langle N_{eh}
\rangle}^{m^{'}}}{m^{'}!}exp(-\langle N_{eh} \rangle),
\end{equation}

\noindent where $\langle N_{eh}\rangle$  is the average number of
excitations inside the nanocrystal. This is a sum of Poissonian distributions
which only depends on the excitation power since the excitation is far above 
the bandgap. For a single nanocrystal weakly pumped, higher order
multiexcitons are not excited. In this case, the probability to emit
two photons through a radiative decay of the biexciton and exciton is $P_{Neh\geq
2}Q_{2X}Q_{X}$, the product of the quantum yields of the two states weighted by the probability 
to excite at least two electron hole pairs, while the probability to emit one photon is
$P_{Neh\geq 1} Q_{X}$. As shown in appendix \ref{sec:App},  $g^{(2)}(0)$
reduces to twice the probability to emit  two photons over
the probability to emit one photon squared:

\begin{equation}\label{g2-Generalized}
g^{(2)}(0)=\frac{2 P_{Neh\geq 2}}{P_{Neh\geq 1}^2}\frac{Q_{2X}Q_X}{Q^2_X}
=\frac{2 P_{Neh\geq 2}}{P_{Neh\geq 1}^2}\frac{Q_{2X}}{Q_X}
\end{equation}

\noindent In a very weak pumping regime such that $\langle
N_{eh}\rangle \rightarrow 0$, Eq.(\ref{g2-Generalized})
simplifies to:

\begin{equation}\label{g2-Nair}
g^{(2)}(0,{\langle N_{eh}\rangle \rightarrow 0})=\frac{Q_{2X}}{Q_X},
\end{equation}

\noindent as $\lim\limits_{\langle N_{eh}\rangle \to 0} (2 P_{Neh\geq
2})/(P_{Neh\geq 1}^2)=1$. This formula has been used in
\cite{nair2011biexciton,park2011near,zhao2012biexciton} to derive
the biexciton quantum yield from an ACF measurement. 
This implies that an emitter having a non zero biexciton quantum
yield $Q_{2X}$ will not show complete antibunching even at very
low pumping regime.

\begin{figure}
  \centering
\includegraphics[scale=0.75]{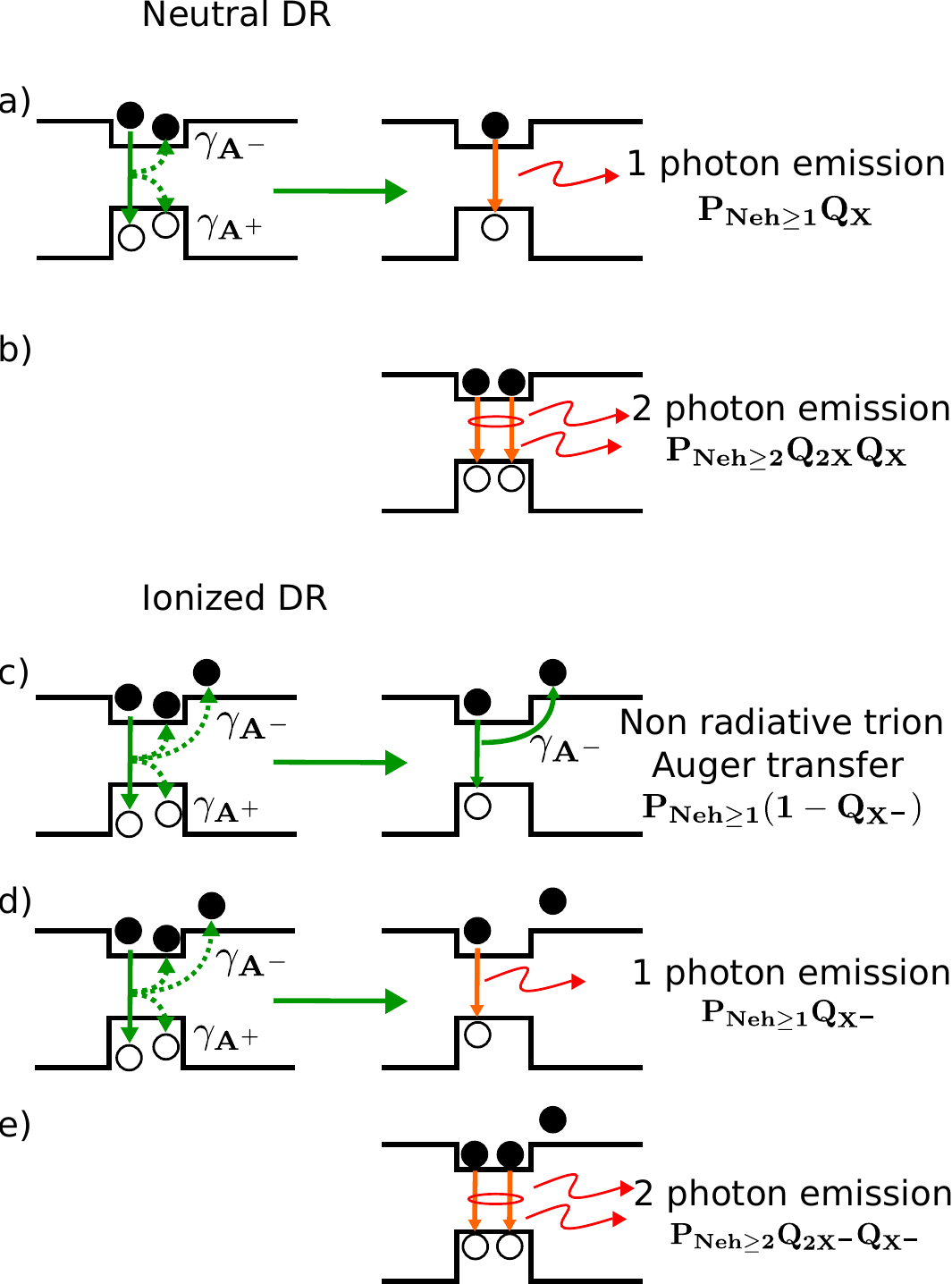}
  \caption{Schematic of the different relaxation pathways after a pulsed excitation of the 
  shell state continuum. Green arows symbolizes the possible Auger relaxation with
  their associated rates $\gamma_{A^{+}}$ if the relaxation energy is given to a hole or $\gamma_{A^{-}}$ if it is
  given to an electron. Dashed green arrows means that different Auger relaxation possibilities are in competition.  
  Orange arrows symbolise radiative recombinations. For the low power excitation considered: $\langle Neh \rangle=0.4$,
  five relaxation cases are possible. 
  For a neutral DR, a) single photon emission after an Auger non radiative decay of the biexciton, b) biexciton
  binding and two photon emission with the given probabilities. For an ionized DR, c) non radiative
  relaxation of the biexciton and exciton, d) single photon emission after Auger non radiative decay of the
  biexciton, e) biexciton binding and two photon emission.}
  \label{fig:Recombinaison}
\end{figure}

In the following we will consider an excitation power such that
$\langle N_{eh} \rangle \sim 0.4$. The 
weighting term in Eq.(\ref{g2-Generalized}) is $(2 P_{Neh\geq 2})/({P_{Neh\geq 1}}^2)=1.13$  in
this case. This excitation power has been chosen
such that the grey state generated by photoionization could be clearly observed. This is
typically realized when the probability to excite at least two electron-hole
pairs is high enough, here $P_{N_{eh} \geq 2}/P_{N_{eh} \geq
1}\simeq 17 \%$. In this case, ionization events through Auger processes
are sufficiently frequent to observe the grey state in the PL intensity
distribution, which is not the case at lower excitation.
At this excitation power $P_{N_{eh} \geq 3}/P_{N_{eh} \geq 1} \simeq 2\%$, this
ensures that higher excited states such as a doubly charged exciton for example are very
unlikely, validating our two states model for the emission. Fig.\ref{fig:Recombinaison} presents 
the different relaxation possibilities for such an excitation,
$\langle N_{eh} \rangle \sim 0.4$.
First possibility, the DR is in a neutral state. Starting from two excitons in the structure, the DR 
will emit one photon with probability $P_{Neh\geq1}Q_{X}$ after Auger non-radiative relaxation of the biexciton
with rate $\gamma_{A^+}$ or $\gamma_{A^-}$ depending to which charge the energy is given to
(a hole or an electron as depicted on Fig.\ref{fig:Recombinaison}a). Or it will emit two
photons with probability $P_{Neh\geq2}Q_{2X}Q_{X}$ if no Auger relaxation takes place (Fig.\ref{fig:Recombinaison}b).
Second possibility, the DR is in a charged state. It will either emit one (Fig.\ref{fig:Recombinaison}d) or two photons
(Fig.\ref{fig:Recombinaison}e) with probabilities $P_{Neh\geq1}Q_{X^-}$ and $P_{Neh\geq2}Q_{2X^-}Q_{X^-}$ respectively,
with $Q_{2X^-}$ the charged biexciton QY. Non radiative Auger transfer to the extra electron with rate $\gamma_{A^-}$ 
can also quench the emission and no photon will be emitted (Fig.\ref{fig:Recombinaison}c).

Quantifying the change in QYs between
neutral and charged DRs using post-selection of 
the photon detection events will allow us to understand the change in photon statistics due to charging. The overall 
photon statistics of a DR resulting in an interplay between the neutral and charged photon statistics  is
explained below.
Differences of photon statistics between DRs is also presented and explained in
section\ref{sec:Charge_delocalization_and_biexciton_emission}.

In the following, Eq.(\ref{g2-Generalized}) will be therefore applied to the
post-selected photons of each state separately, thus providing information
on the biexciton and charged biexciton QYs through the exciton and trion
autocorrelation function:

\begin{equation}\label{g2-XXm}
\left\{
\begin{array}{ccc}
g_{X}^{(2)}(0)& = &\frac{2 P_{Neh\geq 2}}{P_{Neh\geq 1}^2}\frac{Q_{2X}}{Q_{X}},
\\
g_{X^-}^{(2)}(0)& = &\frac{2 P_{Neh\geq 2}}{P_{Neh\geq
1}^2}\frac{Q_{2X^-}}{Q_{X^-}},
\end{array}
\right.
\end{equation}
\noindent To realize this
analysis, the post-selected photons are chosen inside intensity
windows such that the number of photons is maximized to have the
largest statistics and the mix between the two states is minimized.
This last point is assessed using the $g^{(2)}$ values on several
decades of delays between photons as in Fig.\ref{fig:Lifetimes}c. A
$g^{(2)}$ equal to one for the photons from a given intensity window
for all photon delays is an evidence that we are selecting photons 
from only one state of emission.

\begin{figure}
  \centering
\includegraphics[scale=0.5]{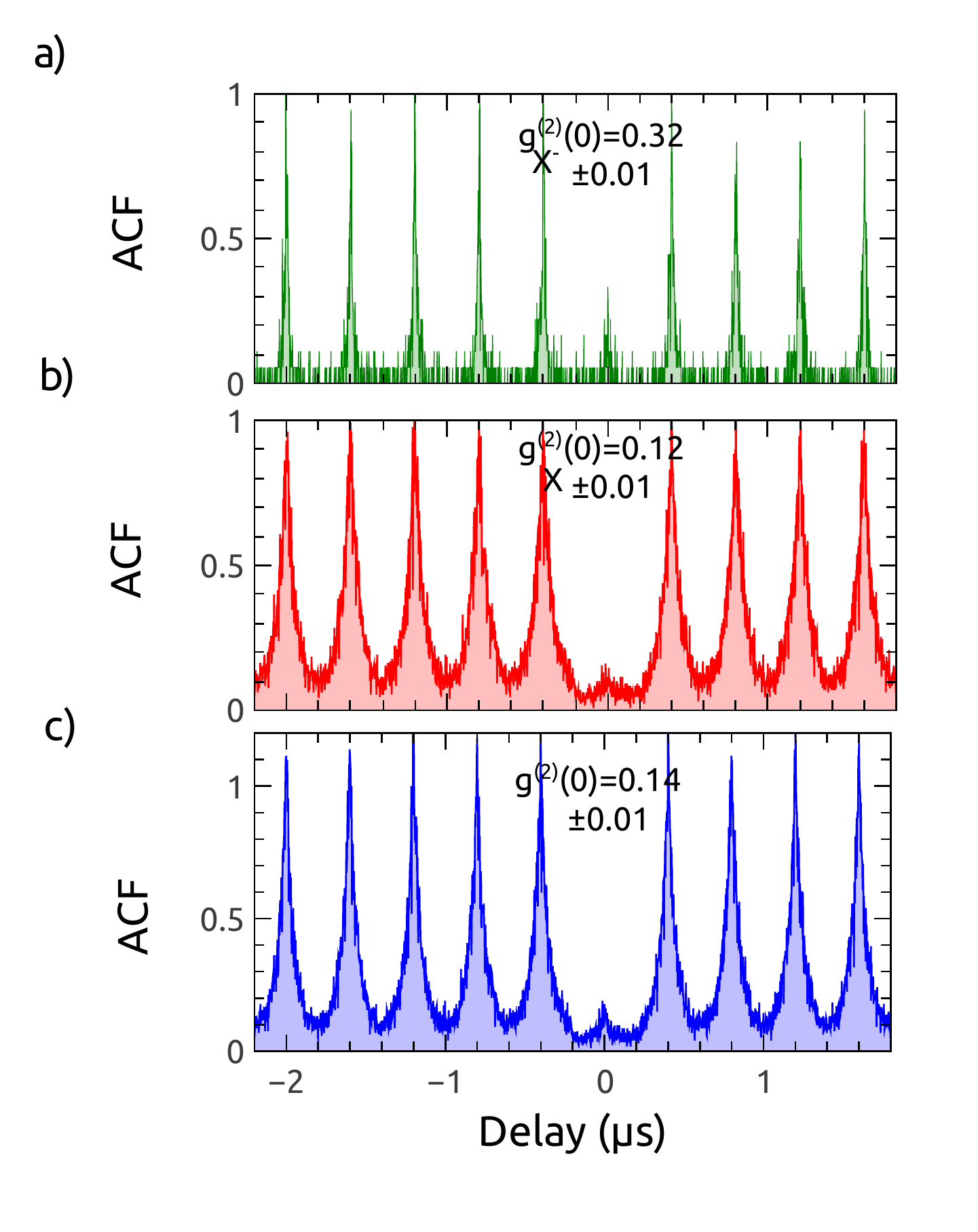}
  \caption{ACF for DR1 at short timescales. \textbf{a)} Grey state ACF.
  \textbf{b)} Bright state ACF.
  \textbf{c)} Whole PL intensity ACF.}
  \label{fig:G2-1}
\end{figure}

Fig.\ref{fig:G2-1} shows the ACF for DR1 whose
emission characteristics were presented in section
\ref{sec:Flickering}. If computed with a time resolution shorter than the emitter lifetime, 
the ACF consists in a series of peaks with a time interval corresponding to laser repetition
rate. In this case, the quantum nature of the emission is evidenced by the antibunching, the peak at zero delay being
smaller than the other peaks. The ACF has been normalized such that to each peak height corresponds the $g^{(2)}$ value
found at longer delays in Fig.\ref{fig:Lifetimes}c. The $g^{(2)}(0)$ value can thus be easily found by looking at the
height of the peak at zero delay.
DR1 gives $g_{X^-}^{(2)}(0) = 0.32$ for the grey state, larger than 
$g_{X}^{(2)}(0)= 0.12$ for the the bright state by a factor of $2.7$. A degradation of the
single photon emission properties is therefore associated with a DR
in a charged state. We can also observe that for all the detection events 
$g^{(2)}(0)=0.14$, it is close to the
bright state photon statistics as $g^{(2)}_{X}(0)=0.12$. The grey
state photon proportion for this measurement is low: $6.5\%$, the
bright state photons representing $76.5\%$ of the measured events
and the discarded photons $17\%$ as already mentioned in section \ref{sec:Flickering}. 
Hence, the bright state photon statistics is very close to the overall photon
statistics.

As can be seen from Eq.(\ref{g2-XXm}), the degradation of the single
photon emission with charging is related to the increase of the
ratio between the biexciton and exciton QY. 
A higher $g^{(2)}(0)$ value is therefore not necessarily due to an increase of the two photon
emission but can also be the result of a decrease in the single
photon emission. In order to go further, the various QYs involved
in Eq.(\ref{g2-XXm}) have to be quantified to see the effect of the charging 
on the single photon and two photon emission. 

In quasi-type II structures such as CdSe/CdS DRs
\cite{pisanello2013}, the conduction band offset between the
two materials being fairly weak the electron is not well confined
inside the CdSe core but instead delocalized on the whole structure.
Conversely, the holes are well confined into the CdSe core due to a
higher offset between the valence bands of the two materials. The
Auger rates to a hole $\gamma_{A^{+}}$ and to an electron $\gamma_{A^{-}}$ can therefore be
different as they scale with the confinement volume
\cite{robel2009universal}. In the following section, we study  the
consequences of different Auger rates corresponding to
energy transfer to positive (hole) and negative (electron) charges
on the photon statistics. To do so we present measurements on another DR
(namely DR2) from the same sample as DR1 that
presents a much shorter exciton lifetime $\tau_X=28$~ns
than DR1 ($\tau_X=65$~ns). As the
value of the exciton lifetime is directly linked to the overlap
between the electron and hole wavefunctions, it is an indication of
the electron delocalization. We expect different
photon statistics for these two DRs.

\section{Charge delocalization and biexciton
emission}\label{sec:Charge_delocalization_and_biexciton_emission}

\begin{figure*}
  \centering
\includegraphics[scale=0.5]{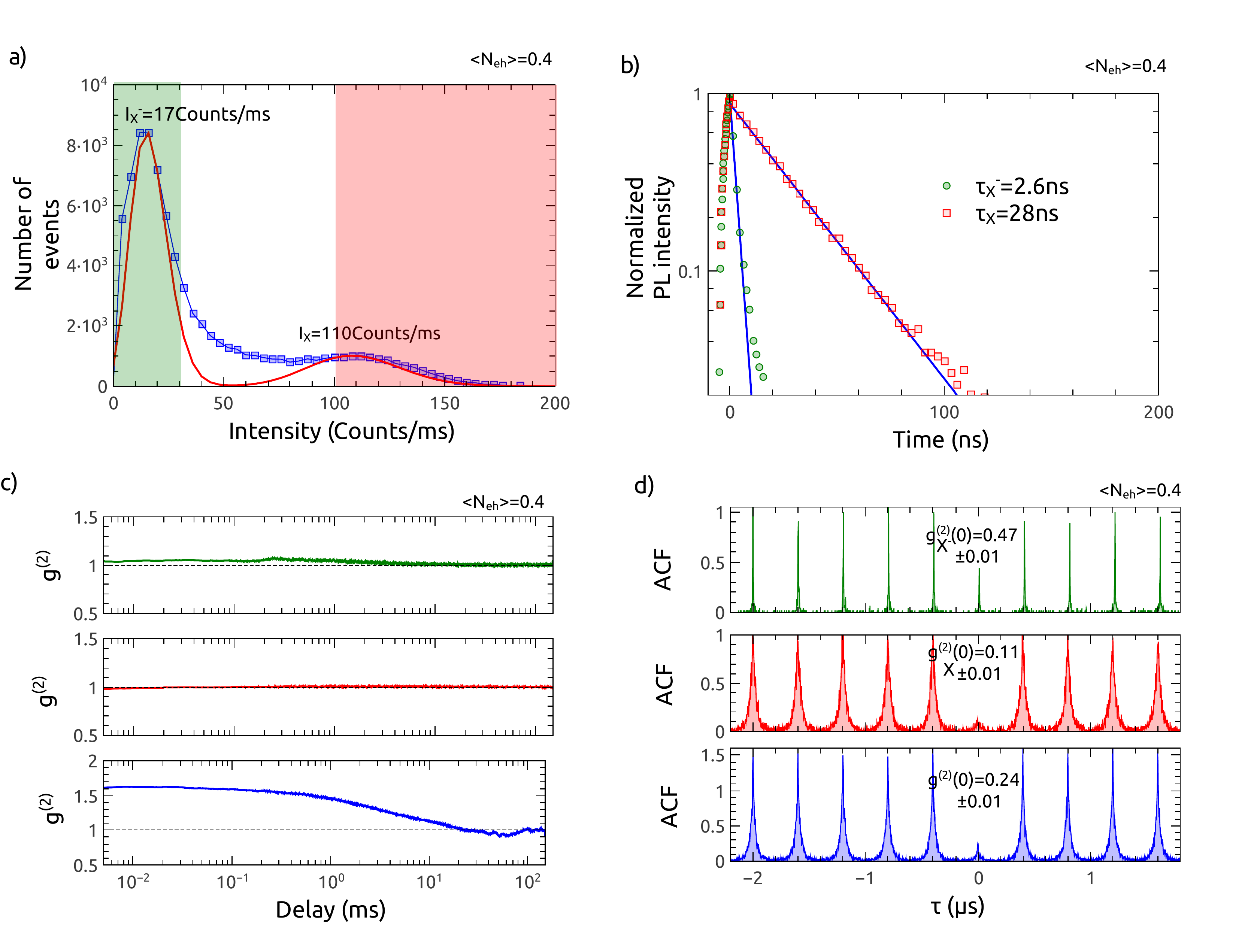}
  \caption{\textbf{a)}Histogram of the PL intensity of DR2 (number of events corresponding to a given
intensity per time bin) for a mean number of excitons in the crystal of $\langle Neh \rangle=0.4$, bin time
$250\mu$s. Count rates below $30$ counts/ms in green are
associated with the grey state, while the part of the histogram above $100$
counts/ms is attributed to the bright state.
  \textbf{b)}PL decay curves for grey state photons (green circles) and bright
state photons (red squares).
  \textbf{c)}$g^{(2)}$ for DR2 on several decades of delays between the
photons. From top to bottom: $g^{(2)}$ for the grey state photons, bright state
photons together with the $g^{(2)}$ of the whole PL intensity.
  \textbf{d)}ACF for DR2 at short timescales. From top to bottom: ACF for the
grey state photons, bright state photons together with the ACF of the whole PL
intensity.}
  \label{fig:G2-2}
\end{figure*}

We now proceed to a comparison between DR1 characterized by a long
lifetime ($\tau_X=65$~ns)  and therefore a highly delocalized
electron and DR2 characterized by a shorter lifetime
($\tau_X=28$~ns) and consequently a more localized electron.
Fig.\ref{fig:G2-2} shows the histogram of PL emission, the PL decay
and photon statistics of DR2. First, it appears that this DR is
characterized by a lower grey state QY, $Q_{X^-}=2\times
2.6/28=18\%$ compared to DR1 ($36\%$).  The trion state is less
emissive in this case because of a higher efficiency of the Auger
relaxation process to the extra electron, the only non-radiative
decay channel. A higher efficiency of the Auger process is
explained in this case by the more confined electron compared to DR1
as the Auger effect scales with
the volume occupied by the charge  \cite{robel2009universal}.
The $g^{(2)}$ of all the photon detection events
(Fig.\ref{fig:G2-2}c bottom) is  characterized by a bunching on the
same timescales as DR1, from microseconds delays to tenth of
milliseconds. The bunching value of $1.6$ is higher owing to a
higher discrepancy of QYs between the two states together with an
increased grey state photon proportion ($18\%$) compared to DR1
($6.5\%$). The post-selection of photon events with
count rates  below $30$ counts/ms and
above $100$ counts/ms for the grey and bright states respectively in
Fig.\ref{fig:G2-2}a allows us to discriminate the photons of each
state. The $g^{(2)}$ corresponding to the bright state
(Fig.\ref{fig:G2-2}c middle) has a value of 1 at all delay
timescales. A limited bunching is still visible for the grey state
$g^{(2)}$ with a value of $1.05$ at short delays. Taking a
smaller intensity window does not change this value. A fast
flickering dynamic between the two states for DR2 that cannot be
resolved correctly with the short binning time of $250\mu$s
very likely explains an imperfect photon
sorting and consequently this small remaining bunching.
For this specific DR we discard $42\%$ of the registered photons.

The bright state photon statistics for DR2 is similar to DR1,with
$g^{(2)}_{X}(0)=0.11$ for DR2 versus $0.12$ for DR1. The two neutral
DRs have comparable single (excitonic) and two photon (biexcitonic)
emission statistics. The grey state photon statistics is
nevertheless different. The degradation of the single photon
emission is more important for DR2 with $g_{X^-}^{(2)}(0)=0.47$
versus $g_{X^-}^{(2)}(0)=0.32$ for DR1. For DR2 the
$g^{(2)}(0)$ value corresponding to all the
photons is equal to $0.24$, it is almost twice the value found for
DR1. This difference can be explained by the proportion of photons
of each state in the overall measurement. The grey state photon
proportion being three times larger
for DR2  than for DR1, $g^{(2)}(0)$ is therefore
increasing towards the charged exciton photon statistics. The
$g^{(2)}$ at zero delay of the entire photon
events detection reflects therefore the interplay between the two
states characterized by different photon
statistics.

In order to go further in this comparison, we calculate the quantum
yields of the various states. The quantum yield of the trion
$Q_{X^{-}}$ is obtained from the ratio between
the grey and bright state lifetimes assuming $Q_{X} \simeq
1$ \cite{brokmann2004measurement,fisher2004emission}. These results
are given in the third column of Table
\ref{tab:Tab1} for the two considered DRs. The quantum yield of the
trion varies from $36\%$ (DR1) to $18\%$
(DR2) of the exciton quantum yield. Then $Q_{2X}$ and $Q_{2X^{-}}$
are calculated using Eq.(\ref{g2-XXm}), with the values of
$g^{(2)}_{X}(0)$ and $g^{(2)}_{X^-}(0)$ from  diagrams in Fig.\ref{fig:G2-1} for DR1 and
Fig.\ref{fig:G2-2}d for DR2 and with $\frac{2 P_{Neh\geq
2}}{P_{Neh\geq 1}^2}=1.13$ corresponding to $ \langle
N_{eh}\rangle=0.4$. Table \ref{tab:Tab1} gives the quantum yield of
the neutral biexciton $Q_{2X}$ and of the
charged biexciton $Q_{2X^{-}}$. The values of the neutral biexciton quantum yields $Q_{2X}$ are
similar for the two DRs and close to $10\%$, corresponding to similar
low values of $g^{(2)}_{X}(0)$ obtained for the neutral state. The
values for the charged biexciton quantum yields $Q_{2X^{-}}$
($6.2\%$ for DR2 and $10.1\%$ for DR1) are found to be from $60$\% to almost $100$\%
of the neutral biexciton quantum yield.
This means that charging affects the biexciton quantum yield less
than the exciton quantum yield. In fact, as mentioned above,
the QY of the charged exciton (trion) is much smaller than the one
of the exciton, yielding an increase of $Q_{2X^{-}}/Q_{X^-}$ compared to $Q_{2X}/Q_{X}$. Thus,
remarkably, the increase of $g^{(2)}(0)$ when a DR is charged is not
due to a higher probability of two photon emission compared to a
neutral DR but to the fact that the two photon emission probability
decreases slower than the single photon emission probability
squared.

Finally, the evolution of the various QYs for these two DRs
allows us to gain information on the relaxation processes in the
DRs. As mentioned above, the dominant non radiative
relaxation process is the Auger effect. The Auger relaxation rate
$\gamma_{A^{-}}$ of an electron hole pair to a neighboring electron
can be derived from the negative trion state QY
\cite{galland2012lifetime}:

\begin{equation}\label{Auger1}
Q_{X^{-}}=\frac{2\gamma_r}{2\gamma_{r}+\gamma_{A^{-}}}.
\end{equation}

\noindent In the case of a biexciton, the energy of an
electron-hole pair can be transferred to a negative or a positive
charge (electron or hole) by an Auger process as depicted in Fig.\ref{fig:Recombinaison}a, c and d. Assuming again that
the non radiative decay channels are due only to the Auger effect
the biexciton QY can be written as \cite{galland2012lifetime}:

\begin{equation}\label{Auger2}
Q_{2X}=\frac{4\gamma_r}{4\gamma_r+2\gamma_{A^{+}}+2\gamma_{A^{-}}}.
\end{equation}

\noindent This formula allows to deduce the Auger relaxation
rate to a positive charge $\gamma_{A^{+}}$ from the quantum yield of
the biexciton. 

For DR1, $\tau_{A^{-}}=1/\gamma_{A^{-}}$ is found to
be $18.3$~ns while the Auger relaxation to positive charges
$\tau_{A^{+}}=1/\gamma_{A^{+}}=4.9$~ns is almost four times faster.
The long Auger relaxation lifetime for electrons  can be explained 
by a highly delocalized electron inside the shell as expected from the long exciton 
lifetime $\tau_{X}=65$~ns. Auger
relaxation to an electron is less efficient than to
a hole because of electron delocalization in quasi
type-II heterostructure like CdSe/CdS DRs
\cite{sitt2009multiexciton}. 
In this case, positive charges
constitute a preferred decay channel and the extra negative charge
does not increase the number of non radiative relaxation paths as it
is an inefficient relaxation solution. The biexciton mainly relaxes
giving its energy to a well confined hole whether the DR is charged
or not, thus in Fig.\ref{fig:Recombinaison} the scenari c and d 
are similar to the scenario a. This explains that for
DR1 the negatively charged biexciton QY is the same as the neutral biexciton QY.

In contrast, for DR2 characterized by a shorter exciton lifetime of $\tau_{X}=28$~ns
and consequently a less delocalized electron compared to DR1, 
using the same calculation, $\tau_{A^{-}}=3.1$~ns and $\tau_{A^{+}}=2.9$~ns
have similar values. It implies in this case a decrease of the
charged biexciton QY compared to the neutral biexciton QY. No decay
channel for the biexciton is favored in this case, so when a DR is
charged, the extra negative charge offers an additional non
radiative decay channel, as can be seen on Fig.\ref{fig:Recombinaison}c and d, which decreases the
charged biexciton QY compared to the neutral biexciton QY. These different behaviors probably come from a slightly different
structure of DR2 as compared to DR1, implying a different localization of the
electrons and the holes in the two DRs as expected from the different exciton lifetimes.
Our measurements thus give access to a full characterization
and interpretation of the physical processes taking place in the DRs
including photon emission, and Auger effects to positive and
negative charges.

\begin{table}
 \caption{Exciton lifetime $\tau_X$, $Q_{X}$ , $Q_{X^{-}}$, $Q_{2X}$,
$Q_{2X^{-}}$ QYs and Auger relaxation time constants for negative ($\tau_{A^{-}}$) 
and positive ($\tau_{A^{+}}$) charges
for the different DRs.}
\begin{tabular}{ c c c c c c c c }
& $\boldsymbol{\tau_X}$ (ns) & $\boldsymbol{Q_{X}}$
&$\boldsymbol{Q_{X^{-}}}$  & $\boldsymbol{Q_{2X}}$ & $\boldsymbol{Q_{2X^{-}}}$ 
& $\boldsymbol{\tau_{A^{-}}}$ (ns) & $\boldsymbol{\tau_{A^{+}}}$ (ns)
\\
  \hline
  \textbf{DR1}& 65 & 100\% & 36\%  & 10.6\%  & 10.1\% & 18.3 & 4.9  \\
  \hline
  \textbf{DR2}& 28 & 100\% & 18\%  & 9.7\%  &  6.2\% & 3.1 &  2.9 \\
  \hline
\end{tabular}
  \label{tab:Tab1}
\end{table}

\newpage
\section{Conclusion}

In conclusion we have shown that
CdSe/CdS DRs emission was characterized by a fast switching between
a bright and a grey state of the emission. Thanks to a
post-selection of the emitted photons we could study independently
the lifetimes and the intensity autocorrelation function of the
bright and grey states. From
this method we deduced the quantum yields of the charged exciton, of
the biexciton and of the charged biexciton. 
In particular we have been able to explain the degradation
of the second order autocorrelation function in charged DRs by the
decrease of the single photon emission probability in the charged
exciton rather than by an increase of the two-photon emission
probability.  By comparing two DRs from the same sample displaying different excitons lifetimes and 
consequently different charges localizations, the effects of the interplay between photon emission and Auger
recombinations were illustrated. We were able to explain the differences between the overall 
photon statistics of these two DRs by quantifying the bright and grey state photons proportions and statistics.
With this analysis we have thus obtained a fully quantitative model of single photon and two-photon emission of
CdSe/CdS dot-in-rods with two states of emission.

\begin{acknowledgements}

The authors gratefully thank Jean-Pierre Hermier
for fruitful discussions.
Financial support from the French
research council ANR, under the project SENOQI,
CNANO Sophopol and
projet ITN-Clermont4

\end{acknowledgements}

\appendix
\section{Derivation of Eq.(\ref{g2-Generalized}).}\label{sec:App}

It has been demonstrated in ref.\cite{nair2011biexciton} that the
autocorrelation function at zero delay can be expressed as a
function of the quantum yield of the different multiexcitonic states $Q_m$ and
the Poissonian probability, assuming the independence of the
different multiexcitonic fluorescence  processes:

\begin{equation}\label{g2-1}
g^{(2)}(0)=\frac{\bigg\langle2\displaystyle\sum_{m>1}^{} P_{Neh\geq m}
\displaystyle\sum_{m'<m} Q_m
Q_{m'}\bigg\rangle_t}{\bigg\langle(\displaystyle\sum_{m\geq1} P_{Neh\geq
m}Q_m)^2\bigg\rangle_t},
\end{equation}
\noindent
where $\langle \rangle_t$ stands for a time average over the
measurement acquisition time if any blinking occurs. Considering an
emitter displaying antibunching, therefore having low multiexciton
QY, the previous equation simplifies to:

\begin{equation}\label{g2-2}
g^{(2)}(0)=\frac{\bigg\langle 2\displaystyle\sum_{m>1}^{} P_{Neh\geq
m}Q_{m}Q_1\bigg\rangle_t}{\bigg\langle (P_{Neh\geq 1} Q_1)^2\bigg\rangle_t},
\end{equation}
\noindent
which can be reduced to

\begin{equation}\label{g2-3}
g^{(2)}(0)=\frac{2P_{Neh\geq 2}}{P_{Neh\geq 1}^2}\frac{\langle Q_2
Q_1\rangle_t}{\langle Q_1^2\rangle_t},
\end{equation}
\noindent
for the typical excitation power considered in this paper $\langle
N_{eh}\rangle \simeq 0.4$. If no blinking occurs or if the photons
are post-selected so as to separate the different states, then we
retrieve Eq.(\ref{g2-Generalized}):

\begin{equation}\label{g2-NonBlinking}
g^{(2)}(0)=\frac{2P_{Neh\geq 2}}{P_{Neh\geq 1}^2}\frac{Q_2}{Q_1}.
\end{equation}
\noindent

\bibliography{BiblioG2Lifetime}

\end{document}